\documentclass[conference]{IEEEtran}
\IEEEoverridecommandlockouts
\usepackage{amsmath, amsfonts, amssymb}
\usepackage{gensymb}
\usepackage{algorithmic}
\usepackage{algorithm}
\usepackage{array}
\usepackage[caption=false,font=scriptsize,labelfont=sf,textfont=sf]{subfig}
\usepackage{tikz}
\usepackage{textcomp}
\usepackage{stfloats}
\usepackage{url}
\usepackage[hidelinks]{hyperref}
\usepackage{orcidlink}
\usepackage{verbatim}
\usepackage{graphicx, color, xcolor}
\usepackage{cite}
\usepackage{tabularx}
\usepackage{multirow}
\usepackage{booktabs} 

\begin{document}


\title{AI-empowered Real-Time Line-of-Sight Identification via Network Digital Twins}
\author{Michele Zhu\textsuperscript{1}, Silvia Mura\textsuperscript{1}, Francesco Linsalata\textsuperscript{1}, Lorenzo Cazzella\textsuperscript{1}, Daminano Badini\textsuperscript{2}, Umberto Spagnolini\textsuperscript{1}\\
\small{\textit{\textsuperscript{1}DEIB, Politecnico di Milano, Milan, Italy}},
\small{\textit{\textsuperscript{2}{Huawei Technologies Italia S.r.l., Segrate}}}\\
\small{Email: \textsuperscript{1}\{name.surname\}@polimi.it}, \textsuperscript{2}damiano.badini@huawei.com}




\maketitle

\begin{abstract}
The identification of Line-of-Sight (LoS) conditions is critical for ensuring reliable high-frequency communication links, which are particularly vulnerable to blockages and rapid channel variations. Network Digital Twins (NDTs) and Ray-Tracing (RT) techniques can significantly automate the large-scale collection and labeling of channel data, tailored to specific wireless environments.
This paper examines the quality of Artificial Intelligence (AI) models trained on data generated by Network Digital Twins. We propose and evaluate training strategies for a general-purpose Deep Learning model, demonstrating superior performance compared to the current state-of-the-art.
In terms of classification accuracy, our approach outperforms the state-of-the-art Deep Learning model by 5\% in very low SNR conditions and by approximately 10\% in medium-to-high SNR scenarios. Additionally, the proposed strategies effectively reduce the input size to the Deep Learning model while preserving its performance.
The computational cost, measured in floating-point operations per second (FLOPs) during inference, is reduced by 98.55\% relative to state-of-the-art solutions, making it ideal for real-time applications.
\end{abstract}

\begin{IEEEkeywords} AI, Deep Learning, Real-Time, Network Digital Twins, 6G.
\end{IEEEkeywords}
\section{Introduction}
\label{sect:introduction}



Network Digital Twins (NDTs) are pivotal in accurately characterizing physical wireless propagation environments, enabling informed decision-making across communication system layers. In the era of sixth-generation (6G) networks, NDTs provide an advanced solution for designing complex radio maps by leveraging high-resolution three-dimensional (3D) models. These digital replicas offer real-time, precise representations of the electromagnetic (EM) environment, facilitating adaptive and efficient communication \cite{zhu2024realtime, Pegu2505:Toward}.

A critical aspect of 6G networks is the ability to distinguish between Line-of-Sight (LoS) and Non-Line-of-Sight (NLoS) conditions of User Equipments (UEs), particularly in high-frequency bands like Frequency Range 2 (FR2) and the newly standardized Frequency Range 3 (FR3). NLoS conditions severely impact coverage and reliability, emphasizing the importance of accurate LoS/NLoS classification for optimizing communication and improving localization performance in uplink 5G New Radio (NR) signals \cite{linsalata2022map, 9364875}.

Existing methods for NLoS identification can be broadly categorized into model-based and data-driven approaches.  
Model-based approaches rely on well-defined environmental models or scenarios where data-driven methods rely on collection of massive amount labeled data. Examples include decision-theoretic frameworks based on range measurements~\cite{686556}, channel statistics such as signal power and delay spread \cite{8510840}, and correlation metrics across time, space, and frequency \cite{8886022}.  
Data-driven approaches use real-world or simulated measurements to train models. Classical Machine Learning (ML) methods like Support Vector Machines (SVMs), Random Forests (RFs), and Gaussian Processes (GPs) utilize high-resolution features estimation techniques for LoS/NLoS classification \cite{8322223}. 
Meanwhile, Deep Learning (DL) techniques are able to process raw channel state information (CSI) or low-resolution features, enabling the model to directly learn complex patterns from channel estimates \cite{10121016, 8027020}. Despite their accuracy, DL methods depend heavily on labor-intensive data collection and CSI labeling, often performed via manual visual inspection—a time-consuming and error-prone process unsuitable for real-time 6G applications \cite{8968748}.

To address these limitations, NDTs emerge as a transformative solution for AI-driven approaches. By leveraging synthetic data, NDTs support the development of site-specific LoS identification models without the need for exhaustive measurements. This paper highlights the unique potential of NDTs to overcome challenges in LoS/NLoS classification, paving the way for efficient and scalable 6G networks \cite{zhu2024realtime, DelMoro}.

\textbf{Contributions}
The main contribution of this work are here summarize.
\begin{itemize}
    \item Introduce an AI solution with a deep learning approach based on CNN feature extraction using channel features input for LoS/NLoS identification within an NDT framework. 
    \item The proposed system is trained on data generated from both NDT and Ray-Tracing (RT) simulations, with performance improvements achieved through downsampling and data augmentation applied to the ResNet34 model.
    Performance is evaluated against the SegNet model from and traditional ML methods, such as SVM and RF. Numerical results show a 5--10\% accuracy improvement in low-SNR scenarios; a 98.55\% reduction in inference FLOPs compared to the SegNet model; downsampling reduced ResNet34 FLOPs by 93.8\%, lowering the total to 6\%.
    \item An ablation study on ResNet34 validates that downsampling and data augmentation are effective general-purpose techniques, applicable to various DL architectures and real time applications.
\end{itemize}

\textbf{Organization}
The remainder of this paper is organized as follows. Section \ref{sec:SystemModel} details the system model and LoS identification methods, while Section \ref{sec:realtime} discusses the real-time application of LoS identification. The numerical results are presented in Section \ref{sec:dataset_and_scenario}, and Section \ref{sec:conclusion} draws the conclusions.

\section{System Model \&  LoS Identification Methods } \label{sec:SystemModel}

\begin{table*}[ht!]
\caption{Features used for UE LoS identification.}
\centering
\begin{tabular}{lll}
\toprule
\textbf{Feature}            & \textbf{Description}                                                                 & \textbf{Formula} \\ \midrule
\textit{Received Signal Strength (RSS)} 
& Total received signal power with unitary transmitted power.
& $P_{RSS,k} = \sum_{p=1}^{\bar{N}} |\bar{\alpha}_{k,p}|$ \\ \midrule
\textit{Maximum Received Power ($P_{max}$)} 
& Maximum power across all paths. 
& $P_{max,k} = \max |\bar{\alpha}_{k,p}|^2$ \\ \midrule
\textit{Root mean squared (RMS) Delay Spread ($\tau_{k,rms}$)} 
& RMS delay spread of Multi Path Components (MPC). 
& $\tau_{k, rms} = \sqrt{\frac{\sum^{\bar{N}}_p |\bar{\alpha}_{k,p}|^2 (\bar{\tau}_{k,p} - \overline{\tau}_k)^2}{\sum^{\bar{N}}_p |\bar{\alpha}_{k,p}|^2}}$ \\ \midrule
\textit{Rise-Time ($\Delta\tau_k$)} 
& Delay between strongest and weakest MPC. 
& $\Delta\tau_k = \max_\tau |\bar{\alpha}_{k,p}| - \min(\bar{\tau}_{k,p})$ \\ \midrule
\textit{Angles Of Arrival Spread  ($\theta_{k,rms}$, $\phi_{k,rms}$)} 
& Measures angular dispersion in azimuth and elevation. 
& $\theta_{k, rms} = \frac{\sum^{\bar{N}}_p |\bar{\alpha}_{k, p}|^2  \bar{\theta}_{k,p}}{\sum^{\bar{N}}_p |\bar{\alpha}_{k, p}|^2}$ \\ \midrule
\textit{ADCPM} 
& Angle-delay-power channel profile. 
& $\mathbf{X}_k \triangleq \mathbb{E}\left[\mathbf{G}_k \odot \mathbf{G}_k^*\right]$ \\ \midrule \bottomrule
\end{tabular}

\label{tab:features}
\end{table*}

This section presents the system model and the definition LoS features, providing a comprehensive overview of the techniques utilized for effective LoS/NLoS identification.

We consider a cellular wireless system, where the Base Station (BS) is equipped with a Uniform Planar Array (UPA) with $N \times M$ antennas and operates at carrier frequency $f_c$ with a bandwidth $B$. In this setup, the User Equipment (UE) communicates with the BS in the uplink direction with an omnidirectional antenna. 
The uplink signals are transmitted using an Orthogonal Frequency Division Multiplexing (OFDM) waveform with sampling interval $T_s$, $N_c$ subcarriers, and symbol duration $T_c = N_c T_s$. Each sub-carrier frequency is defined as $f_l = \frac{l}{T_c}$, $l = 0, ..., N_c -1$, while the duration of the Cyclic Prefix (CP) is $T_g = N_g T_s$ where $N_g$ are the allocated subcarries. The CP duration is assumed to exceed the maximum delay experienced by the UEs in the cell: $\max_{k,p} \tau_{k, p}$, where $\tau_{k, p}$ represents the delay of the $p$th path for the $k$th UE. $N_p$ denotes the number of multipath components encountered in the communication.

The Channel Frequency Response (CFR) between the $k$th UE and the BS at the $l$th subcarrier is modeled as
\begin{equation} \label{eq:true_channel}
    \mathbf{h}_{k} [l] = \sum^{N_p}_{p=1} \alpha_{k, p}[l] \mathbf{e} (\theta_{k, p}, \varphi_{k,p}) \in \mathbb{C}^{NM \times 1}.
\end{equation}
where $\alpha_{k,p}[l] = a_{k,p} e^{-j2\pi  f_l \tau_{k,p}}$ is the complex path gain, with $a_{k,p}$ denoting the path amplitude. The vector $\mathbf{e}(\theta_{k,p}, \varphi_{k,p})$ represents the array response \cite{9364875}. 

The received signal $\mathbf{r}_{k}[l]  \in \mathbb{C}^{NM \times 1}$ is
\begin{equation}\label{eq: AWGN}
    \mathbf{r}_{k}[l] = \mathbf{h}_{k}[l] {s}_{k}[l] + \mathbf{z}_{k}[l],
\end{equation}
where ${s}_{k}[l]$ represents the transmitted symbol data such that $\mathbb{E}[{s}_{k}[l] {s}_{k}[l']^*] = \sigma_s^2$ with $l \neq l'$ and $\mathbf{z}_{k}[l] \in \mathbb{C}^{NM \times 1}\sim \mathcal{CN}(0, \sigma_{z}^2\, \mathbf{I}) $ is Addictive Gaussian White Noise (AWGN). 

Multiple features, both high-resolution and low-resolution, can support LoS/NLoS classification. High-resolution features are derived from propagation path estimation using algorithms like MUSIC \cite{doi:https://doi.org/10.1002/9781119294016.ch7}, with key features summarized in Tab. \ref{tab:features}. In contrast, the main feature used in this work is the low-resolution Angle Delay Channel Power Matrix (ADCPM) ($\mathbf{X}_k$), representing the channel's angle-delay-power profile. 
It is defined as
\begin{equation}\label{eq:adcpm}
    \mathbf{X}_k \triangleq \mathbb{E}\left[\mathbf{G}_k \odot \mathbf{G}_k^*\right] \in \mathbb{R}^{MN\times N_c},
\end{equation}
where $\mathbf{G}_k$ is the angle-delay channel response matrix
\begin{equation}
    \mathbf{G}_k \triangleq \frac{1}{\sqrt{MNN_c}} (\mathbf{V}_M \otimes \mathbf{V}_N)^H \mathbf{H}_k \mathbf{F}^*,
\end{equation}
with $[\mathbf{V}_M]_{i,j} = \frac{1}{\sqrt{M}} e^{-j2\pi \frac{i(j-M/2)}{M}}$, $[\mathbf{V}_N]_{i,j} = \frac{1}{\sqrt{N}} e^{-j2\pi \frac{i(j-N/2)}{N}}$, and $[\mathbf{F}]_{i,j} = \frac{1}{\sqrt{N_c}}e^{-j2\pi \frac{ij}{N_c}}$ representing the DFT matrices, while $\mathbf{H}_k  = [\mathbf{h}_k[1], ...,\mathbf{h}_k [N_c]] \in \mathbb{C}^{NM \times N_c}$ is the channel matrix. As discussed later, the ADCPM offers a sparse representation of the communication channel within the delay-angle-power domain that can be used in DL for LoS identification purposes.

\subsection{Traditional ML Methods.} 

The LoS identification task is framed as a supervised binary classification problem, employing both traditional ML methods (SVM, RF) and DL models.
By using the high-resolution feature vector 
\begin{equation}\label{eq:feature_ml}
\mathbf{x}_k = \left[P_{RSS,k}, P_{max,k}, \tau_{k,rms}, \Delta\tau_k, \theta_{k,rms}, \phi_{k,rms}\right]^T.
\end{equation}
two main ML classifiers are used as benchmarks:

\begin{itemize}
    \item \textit{Support Vector Machine (SVM)}: a model to separate classes $\{-1, 1\}$ via a hyperplane, using a linear kernel $k(\mathbf{x}, \mathbf{x}_k) = \mathbf{x}^T \mathbf{x}_k$ or an RBF kernel $k(\mathbf{x}, \mathbf{x}_k) = \mathrm{exp}(-\gamma||\mathbf{x}-\mathbf{x}_k||^2_2)$.

\item  \textit{Random Forest (RF)}: An ensemble model utilizing bagging and decision trees with Gini index-based splits, with $Gini(\mathbf{x}) = 1 - \sum_{z=1}^Z p_z^2$ and $p_z$ is the class ratio.
\end{itemize}

\subsection{Deep Learning (DL) Methods.}
DL convolutional architecture automates feature extraction process using architectures like SegNet \cite{10274097} and ResNet \cite{He_2016}:
\begin{itemize}
   \item  \textit{SegNet:} An auto-encoder that combines input reconstruction and classification. The loss function includes reconstruction loss and binary cross-entropy:
\begin{align}
    \mathcal{L} &(\mathbf{X}_k, y_k) = \frac{w_{rec}}{K} \sum^{K}_{k=1} ||\mathbf{X}_k - \mathbf{\hat{X}}_k||^2_2 \nonumber \\
    &+ \frac{1}{K}\sum^{K}_{k=1} -\left\{ y_k \log(\hat{p}_{y_k}) + (1-y_k) \log(1-\hat{p}_{y_k})\right\},
\end{align}
where $\mathbf{X}_k$ is defined in \eqref{eq:adcpm}, and $y_k \in \{0, 1\}$ represents the LoS/NLoS label with probability $\hat{p}_{y_k}$, ${w}_{rec}$ is a weighing fact for tuning the reconstruction, and $K$ is the number of potential UEs used in the loss function computation.

 \item \textit{ResNet:} A general-purpose DL model optimized solely for classification. Its loss function is:
\begin{align}
    \hspace{-0.6cm}\mathcal{L}(\mathbf{X}_k, y_k) &= 
    \frac{1}{K}\sum^{K}_{k=1} -\{ y_k \log(\hat{p}_{y_k}) 
    + (1-y_k) \log(1-\hat{p}_{y_k})\}.
\end{align}

\end{itemize}
\noindent While SegNet jointly optimizes reconstruction and classification, ResNet offers a simpler alternative with faster training and inference. Both approaches leverage strategies to enhance accuracy, robustness, and generalization on noisy real-world data.
ResNet is a type of DL architecture for image recognition tasks; it prevents the vanishing gradient effect by introducing skip connections. The design principle applied to ADCPM inputs is a CNN feature extraction that learns to identify the key features or patterns in the input samples. These features are fed to a Multilayer Perceptron Layer (MPL) classification head transforming the extracted features into the final prediction label. We will adopt ResNet34 architecture and compare our results with the SegNet in \cite{10274097} as a DL benchmark

\section{Toward real-time DL LoS Identification}\label{sec:realtime}

\begin{figure} [!t]
    \centering
    {\frame{\includegraphics[width = 0.90\columnwidth]{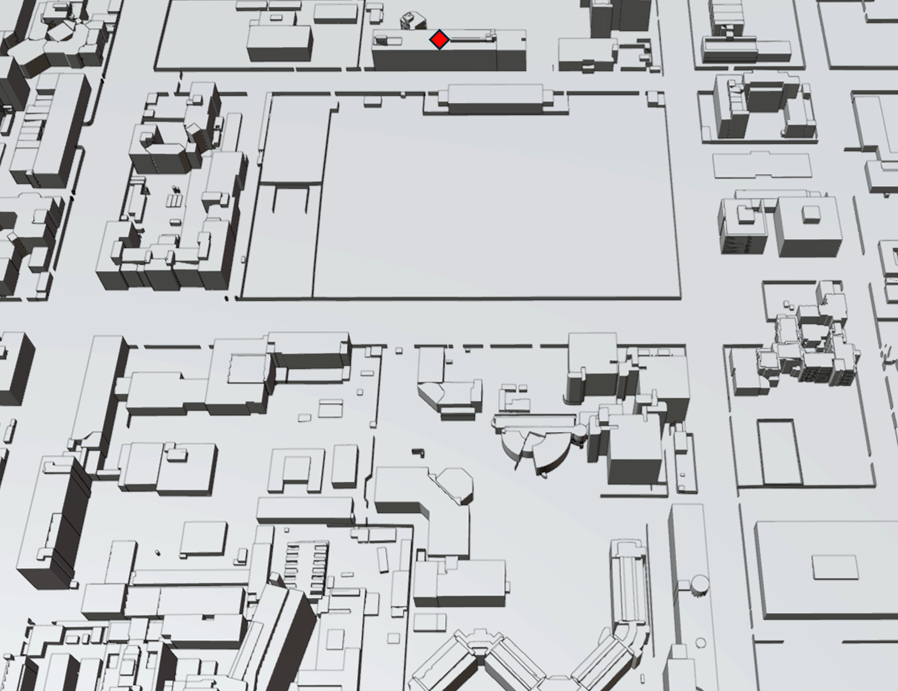}\label{subfig: digital-map}}}
    
    \caption{3D Digital Twin of the urban area of Milan. The red diamond indicates the position of the Base Station.}
    \label{fig: NDT-milan}
\end{figure}

This section provides a comprehensive overview of the essential elements required for achieving real-time LoS/NLoS classification. First, it defines the NDT framework, emphasizing that obtaining reliable and rapidly generated synthetic data is critical for handling real-time applications. Next, the proposed DL architecture is discussed, highlighting the key components that need optimization within this framework to facilitate real-time LoS identification.

\subsection{Network Digital Twin Framework}

The NDT serves as a high-fidelity digital representation of the physical environment, created by integrating accurate urban 3D meshes with ray tracer simulations. In the context of communication, the ultimate goal of the NDT is to accurately detect and model the communication link status of UEs\cite{zhu2024realtime}.

To enhance its decision-making accuracy and improve its approximation of the physical world, the NDT must undergo continuous refinement. Dynamic scenarios, such as those encountered in 6G networks, are often characterized by stringent time requirements and tolerance constraints. To meet the demands of these applications, the NDT must be updated quickly and reliably.

The proposed NDT simulation framework facilitates the generation of realistic datasets comprising synchronized positional and wireless channel data. Channel propagation is simulated using the Sionna Fibonacci ray tracer \cite{Hoydis_2023} along with an accurate 3D digital map of the environment, as shown in Fig. \ref{fig: NDT-milan}. 
This digitized replica of the real world generates channel data that is utilized to train classification models. The trained models are then employed in real-world scenarios to classify the estimated channel in the uplink. 

It is essential to emphasize that uplink signals and communication channel measurements can be utilized to update and improve the accuracy of the NDT representation, as well as to further refine the classification models.

\subsection{Deep Learning for real-time classification}

\begin{figure} [!t]
    \centering
    \subfloat[Sample in LoS]{\includegraphics[width=0.45\columnwidth]{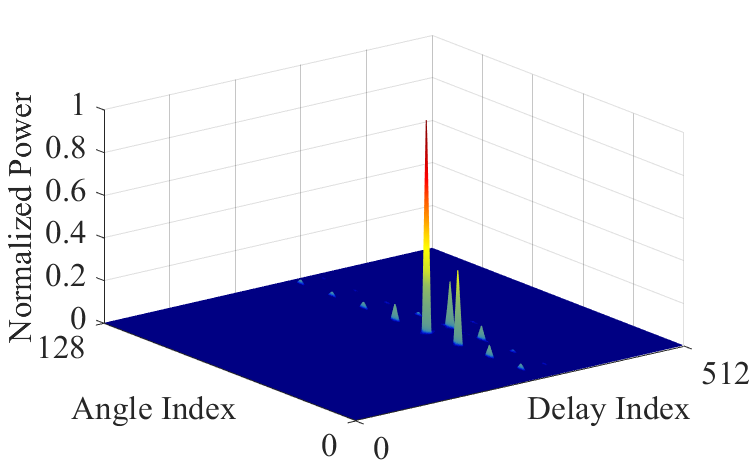}\label{subfig: adcpm_los}}
    \subfloat[Sample in NLoS]{\includegraphics[width=0.45\columnwidth]{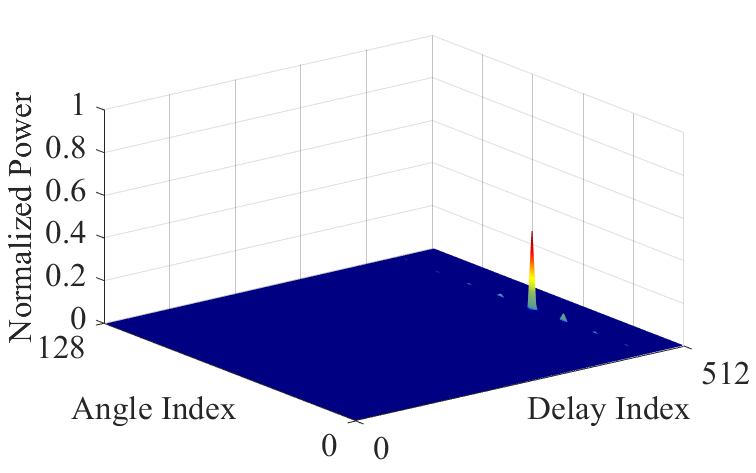}\label{subfig: adcpm_nlos}}
    
    \subfloat[Max pooling]{\includegraphics[width=0.45\columnwidth]{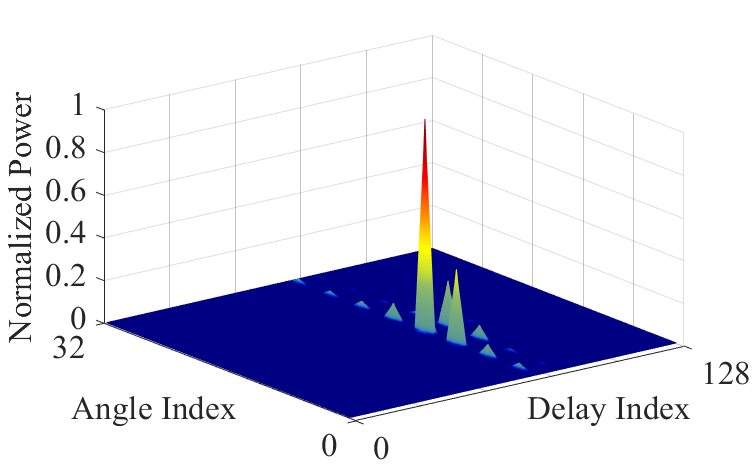}\label{subfig: adcpm_downsampled}}
    
    \caption{Angle-Delay-Channel-Power-Matrix samples in LoS (a) and NLoS (b) condition with normalized power. (c) is the sample in (a) after max pooling operation with a kernel(4, 4), the overall structure is not impacted by the operation.}
    \label{fig: ADCPM-sample}
\end{figure} 

In this subsection, we introduce the proposed DL architecture, highlighting its design and characteristics, and we analyze the proposed methods to improve model performance in computational time and classification accuracy.

The analyzed DL models, which take ADCPM as input, have a computational cost dependent on $(MN, N_c)$—the number of antennas in the array and subcarriers, respectively. However, deploying large antenna arrays in 6G Massive MIMO systems and using a high number of subcarriers can lead to prohibitive computational and storage demands for real-time applications \cite{CC-complexity}.

This issue is mitigated by first utilizing ResNet to enable faster training and inference, and then by thoroughly analyzing the input matrices. As illustrated in Fig.~\ref{subfig: adcpm_los}, the ADCPM is highly sparse, and much of the angular and temporal resolution in this representation is unnecessary for classification tasks. Consequently, it is feasible to reduce the dimensions of the network inputs $\mathbf{X}_k$ without losing critical information. This reduction is achieved by downsampling the input using a max-pooling operation, which also preserves the relative locality of the MPC in the ADCPM as shown in Fig.~\ref{subfig: adcpm_downsampled}.
In this process, non-overlapping rectangular windows slide over the input matrix, and the output consists of the maximum value within each window. This operation improves the real-time feasibility of the analyzed DL models.

Moreover, downsampling can be used to manage the varying number of subcarriers, which can become an additional hyper-parameter in a wireless communication system. 
DL models, in particular the fully connected classifier, are trained to handle a fixed dimension. The max-pooling approach can be used to downsample the subcarrier dimension to a fixed width before feeding it to the model itself.

Overfitting is another significant challenge in which a model accurately classifies training data points but struggles to generalize to new unseen data. In particular, wireless are highly unpredictable, where factors such as user mobility and environmental condition can drastically change the wireless channel. Data augmentation is often used in the computer vision community to improve deep learning models and mitigate overfitting. To address this issue, we propose to use the noisy NDT-powered channel matrix, obtained by adding AWGN to $\mathbf{\tilde{X}}_k$, to train the DL model.

\section{Numerical Results} \label{sec:dataset_and_scenario}

This section reports the main numerical results of the classification algorithms in an urban scenario of the city of Milan, Italy.

\subsection{Training and testing settings}

The simulations are performed with the carrier frequency $f_c = 28$ GHz, bandwidth $B=400$ MHz, and $N_c=512$. A single BS is placed at the top of one building at a height of $21.7$ m. The BS is equipped with an isotropic antenna array with $N=8$, $M=16$ rows and columns, the antenna array has a spacing of $0.8 \lambda$ and $0.5 \lambda$ in the vertical and horizontal directions. The antenna follows a directional pattern as specified in 3GPP TR 38.901 \cite{3gpp_antenna} and is oriented at $135^\degree$ relative to the north direction.
The scenario in Fig.~\ref{subfig: digital-map} is divided into squares of $2\times2  \text{m}^2$, forming a grid of $7.5 \cdot 10^4$ potential UEs positions, corresponding to the center of the grid samples. 
We randomly sample $6 \cdot 10^3$ positions for the test set and remove them from the train set. 

Direct training and evaluation of the proposed model using the NDT data only would lead to overly optimistic results, often resulting in poor performance when applied to real-world measurements. To address this, testing is conducted based on the estimated communication channel $\hat{\mathbf{H}}_k$ obtained through the 5G NR standard-compliant channel estimation method outlined in \cite{hassan2020channel}, derived from the uplink signal in \eqref{eq: AWGN}.

We employ k-fold cross-validation for the ML approaches~\cite{10.5555/1162264}, where training utilizes NDT-generated high-resolution features, and testing is conducted on the same features outlined in Sect.~\ref{sec:SystemModel}, derived from 5G NR uplink channel MPC estimates. For the proposed DL approach, we use a hold-out evaluation, with the training set consisting of NDT-generated ADPCM matrices enhanced with AWGN noise to ensure a fixed SNR across all samples. This data augmentation acts as a regularization factor to mitigate overfitting of the DL model, as discussed in the previous section. Testing is performed on the 5G NR uplink ADPCM estimates.

\subsection{Accuracy, ROC analysis and ablation study} \label{subsec: analysis}

Accuracy is defined as the ratio of correctly classified samples to the total number of samples. For the accuracy plots, we apply a classification threshold of 0.5 to the probabilistic model output. For binary classification models that output probabilities, performance across various classification thresholds is assessed using the receiver operating characteristic (ROC) curve. The ROC curve is generated by plotting the True Positive Rate (TPR) against the False Positive Rate (FPR) for every possible threshold. An ideal model achieves a TPR of 1.0 and a FPR of 0.0 at some threshold, while a random classifier on a perfectly balanced dataset will produce an ROC that follows the diagonal from (0,0) to (1,1). The Area Under the Curve (AUC) of the ROC quantifies the model's ability to distinguish between classes, a perfect binary classifier has a AUC score of 1.0. The ROC curve can also guide the selection of an optimal classification threshold depending on the relative importance of TPR or FPR. However, ROC is most accurate when the dataset is balanced; in our case, approximately 40\% of the UE positions are under LoS conditions.

\begin{figure} [!t]
    \centering
    \subfloat[Accuracy vs test dataset SNR]{\includegraphics[width =0.8\columnwidth]{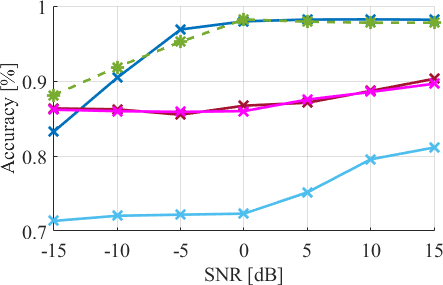}\label{subfig: acc-all}}
    
    \subfloat[ROC for test dataset at -15 dB]{\includegraphics[width =0.8\columnwidth]{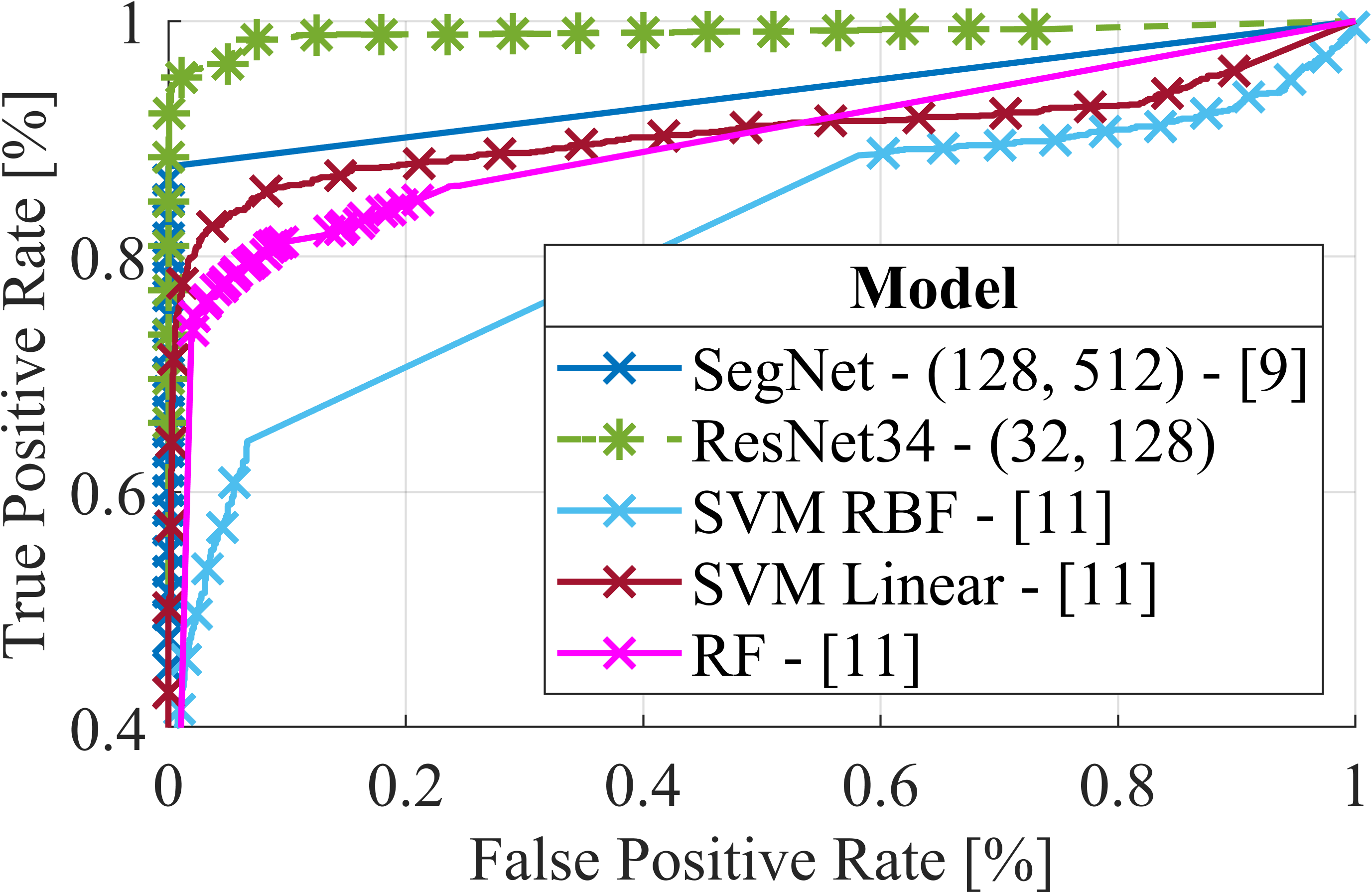}\label{subfig: roc-all}}    
    \caption{Model performance against test set. On the left \protect\subref{subfig: acc-all} Accuracy Score, on the right \protect\subref{subfig: roc-all} ROC curve at -15 dB.}
    \label{fig: comparison-all}
\end{figure}

The results of the LoS identification accuracy are illustrated in Fig.~\ref{subfig: acc-all}. The SVM and RF models show a significant performance drop caused by the discrepancies in the NDT data and the MPC estimated from Eq.~\eqref{eq: AWGN}.
In contrast, ResNet34 and SegNet demonstrate greater robustness, the learned features of $\mathbf{X}_k$ by the CNN feature extractor are more robust than the hand-crafted ones, and they do not require an explicit MPC estimation. We can observe that the two DL models have similar performance when the SNR is higher than 0 dB, in the low SNR regime the SegNet model performance declines faster than our proposed ResNet34 with downsampling and data augmentation. 
In Fig.~\ref{subfig: roc-all}, we show the ROC curve at $-15$ dB, it can be observed that the proposed model has a better generalization performance and is less sensitive to noise. In fact, the SegNet model cannot distinguish the two classes correctly. 

In Table~\ref{tab: auc-all} the AUC is reported for SNR levels of -15, 0, 15 dB. The AUC quantifies the ability of the proposed model to distinguish between classes and aids in the selection of a classification threshold for the model, the proposed ResNet34 (32, 128) outperforms all other classifiers in the AUC score. The high AUC score indicates that the model's accuracy can be further enhanced by selecting a classification threshold more suitable than 0.5.

\begin{figure} [!t]
    \centering
    \subfloat[Accuracy vs test dataset SNR]{\includegraphics[width =0.8\columnwidth]{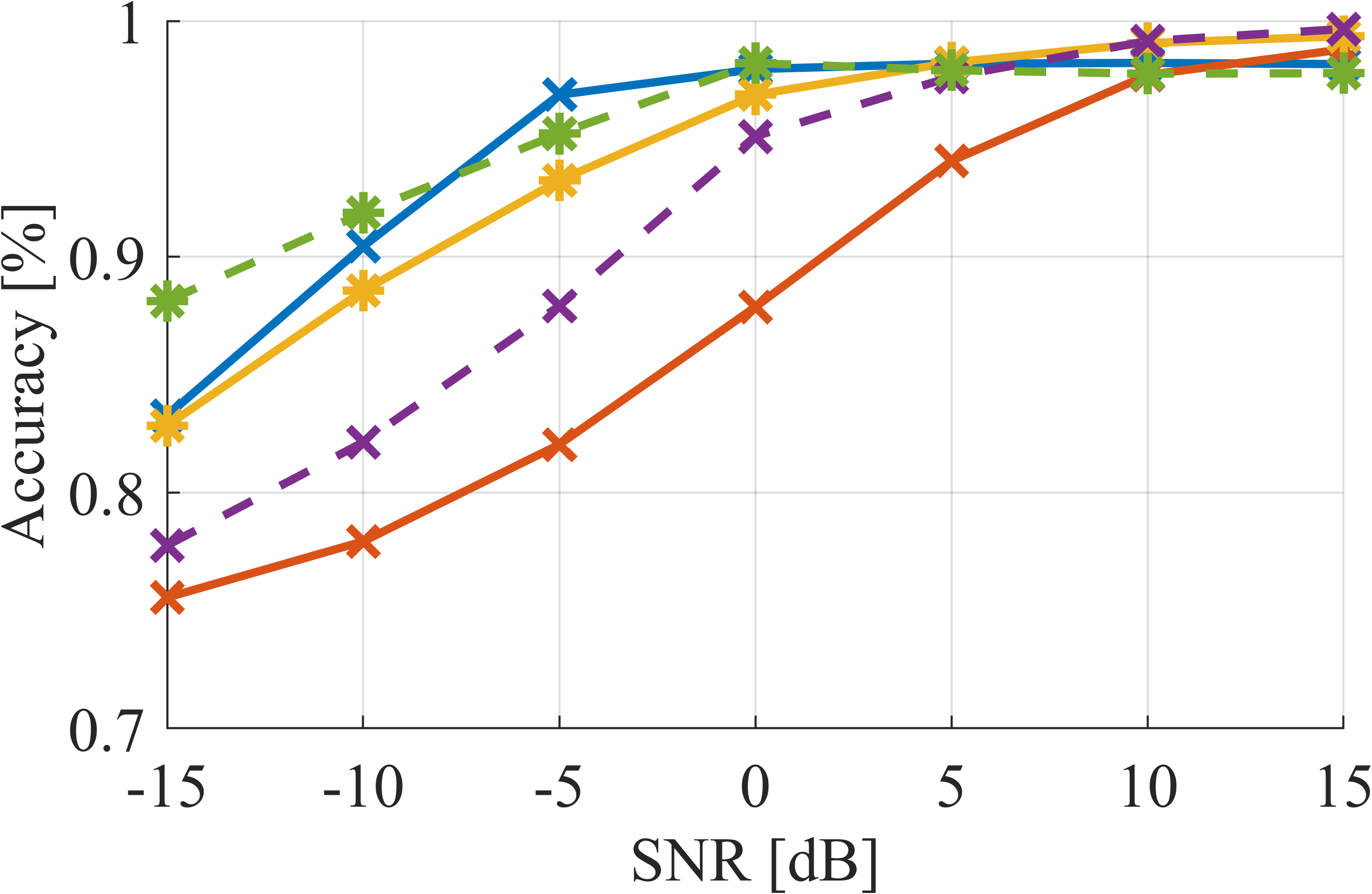}\label{subfig: ablation}} 
    
    \subfloat[ROC for test dataset at -15 dB]{\includegraphics[width =0.8\columnwidth]{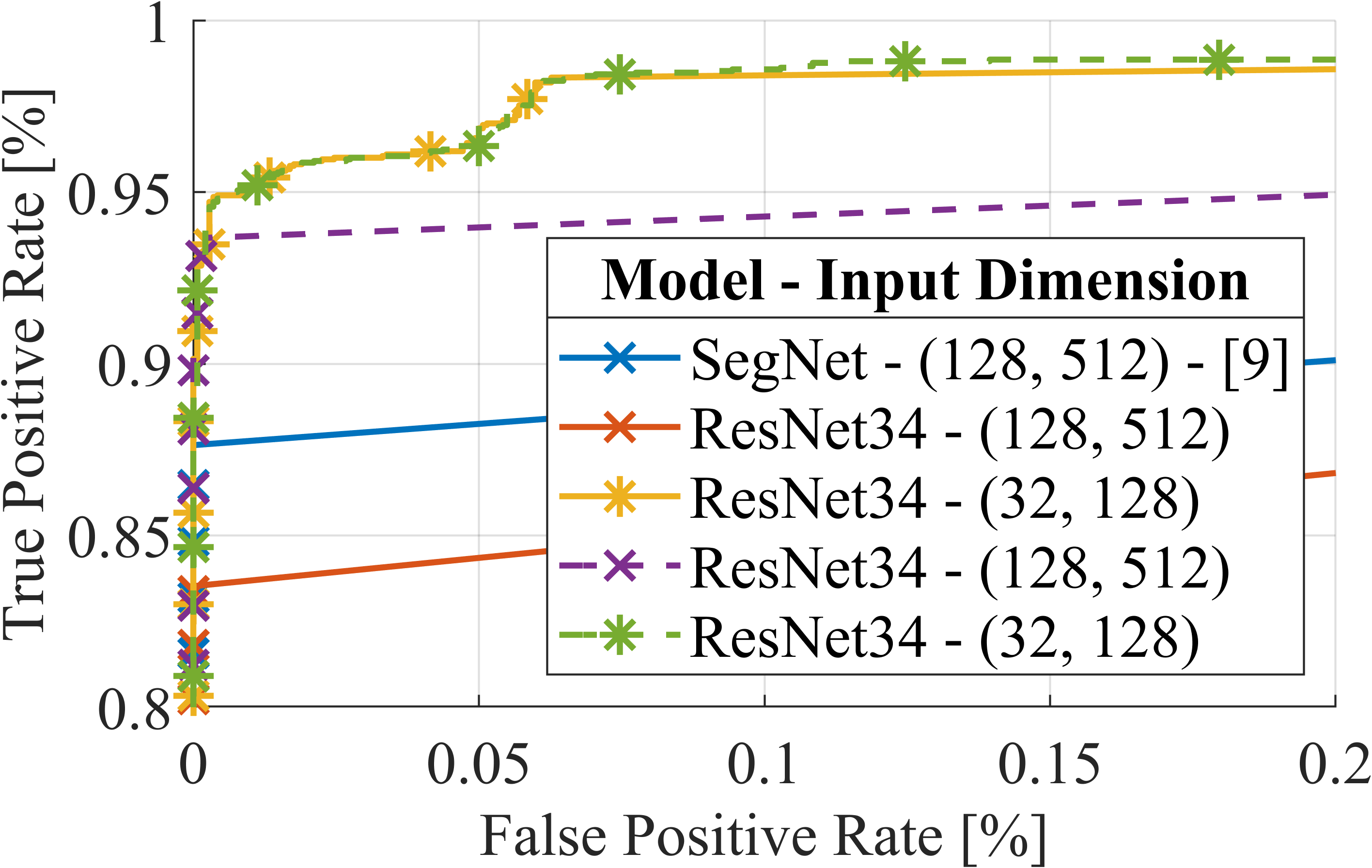}\label{subfig: roc-dl}}   
    \caption{Ablation study against test set. On the left \protect\subref{subfig: ablation} Accuracy Score, on the right \protect\subref{subfig: roc-dl} ROC curve at -15 dB. The dashed lines represents models with data augmentation, solid lines without data augmentation. Input dimension of (32, 128) is the dimension after downsampling.}
    \label{fig: DL}
\end{figure}

To demonstrate the effectiveness of the proposed data augmentation and dimension reduction techniques, we conduct an ablation study shown in Fig. \ref{fig: DL}. In Fig. \ref{subfig: ablation}, the accuracy of ResNet34 trained without these techniques (solid red line) is slightly better than SegNet (solid blue line) only in the high-SNR regime, but its performance drops significantly as the SNR decreases. Applying data augmentation (dashed purple line) slightly improves accuracy, while downsampling (solid yellow line) shows more noticeable improvements. The combination of both techniques (green dashed line) results in superior performance compared to the baseline (solid blue line). Fig.~\ref{subfig: roc-dl} illustrates the model's ability to distinguish between classes, where data augmentation proves to have the most significant impact.

We apply downsampling using max-pooling operations with a window size of $(4, 4)$, which incurs a negligible computational cost compared to the overall cost of the model. SegNet operates as an encoder-decoder architecture, where classification is performed on the latent representation, and during inference, only the encoder is utilized. The computational cost in FLOPs for DL models is summarized in Table~\ref{tab: dl-cost}~\cite{7299173}. The downsampling technique not only enhances performance in low-SNR conditions but also achieves this with a fraction of the original computational cost. The applied max-pooling with kernel $(4,4)$ improves ResNet34 inference cost by 93.8\% against the original cost and 98.55\% against SegNet in \cite{10274097}.  

\begin{table}[!t]
    \centering
    \footnotesize
    \caption{Area Under the ROC (AUC) for different quantizations of SNR.
    }
    \begin{tabular}{ l c c c}
        \toprule
         \multicolumn{1}{l}{Model} &
         \multicolumn{3}{c}{AUC} \\
         \cmidrule(lr){2-4}
         &
         \multicolumn{1}{c}{-15 dB} & \multicolumn{1}{c}{0 dB} & \multicolumn{1}{c}{15 dB} \\ \midrule

         SegNet & 0.9380 & 0.9941 & 0.9666 \\
         ResNet - (32, 128) & 0.9899 & 0.9984 & 0.9984 \\
         SVM - RB & 0.8111 & 0.8247 & 0.8859 \\
         SVM - Linear & 0.9049 & 0.8960 & 0.9458 \\
         RF & 0.8935 & 0.8916  & 0.9127 \\

         \bottomrule
    \end{tabular}
    \label{tab: auc-all}
\end{table}

\begin{table} [!t]
    \centering
    \footnotesize
    \caption{DL model inference computational cost. SegNet is an encoder-decoder architecture and only the encoder is used during inference time.} 
    \begin{tabular}{l c c}
		\toprule
		\textbf{Model - Input Size}  & \textbf{Total Computational Cost} & \textbf{Encoder Cost}\\ 
		\noalign{\smallskip}
		\hline
		\noalign{\smallskip}
            SegNet - (128, 512) & 80 [GFLOPS] & 40 [GFLOPS] \\ \noalign{\smallskip}
            ResNet34 - (128, 512) & 9.36 [GFLOPS] & - \\\noalign{\smallskip}
            ResNet34 - (32, 128) & 0.58 [GFLOPS] & - \\\noalign{\smallskip}
	\bottomrule
    \end{tabular}
    \label{tab: dl-cost}
\end{table}

\section{Conclusion}\label{sec:conclusion} 

DL models often require extensive measurement campaigns to generate training samples, along with labeled data for supervised learning. By leveraging the high-accuracy digital representation and propagation simulation of NDT systems, fast and automated data collection at the micro-cellular level becomes feasible, enabling the creation of site-specific classification models.
This paper demonstrates the effectiveness of NDT-powered data collection for training data-driven classification models. A novel DL classification approach is proposed, combining joint data augmentation and downsampling techniques to reduce input size, computational costs, and enhance LoS/NLoS classification accuracy. The proposed ResNet34-based model achieves \textbf{5\% higher accuracy} than the baseline in very low SNR regimes, \textbf{8\% higher accuracy} than SVM and RF in medium-to-high SNR conditions, and \textbf{93.8\% reduction} in inference cost through downsampling, with a \textbf{98.55\% improvement} over other evaluated DL-based models, making it suitable for real-time applications.

\bibliography{bibliography}
\bibliographystyle{IEEEtran}

\end{document}